\documentclass[12pt]{iopart}

\usepackage{amssymb}  
\usepackage{harvard}

\def\ack{\section*{Acknowledgements}}

\usepackage[pdftex]{graphicx}
\graphicspath{{./Figs/}}

\def\parX{\ensuremath{a}}
\def\parY{\ensuremath{b}}

\def\ie{\emph{i.e., }}
\def\cf{\emph{cf. }}

\def\rmd{\ensuremath{\textrm{d}}}

\newcommand\Sect[1]{Section~\ref{sec:#1}}
\newcommand\sect[1]{section~\ref{sec:#1}}
\newcommand\sects[1]{sections~\ref{sec:#1}}
\newcommand\FIG[1]{Figure~\ref{fig:#1}}
\newcommand\Fig[1]{figure~\ref{fig:#1}}
\newcommand\Figs[1]{figures~\ref{fig:#1}}
\newcommand\fig[1]{\ref{fig:#1}}
\newcommand\Eq[1]{equation~(\ref{eq:#1})}
\newcommand\eq[1]{(\ref{eq:#1})}

\begin{document}

\title[Basin boundary, edge of chaos, and edge state in a two-dimensional model]%
{Basin boundary, edge of chaos, and edge state\\in a two-dimensional model}

\author{J\"urgen Vollmer$^{1,2}$, Tobias M. Schneider$^{2}$, and Bruno Eckhardt$^{2}$}

\address{
$^{1}$  Max-Planck-Institut f\"ur Dynamik und Selbstorganisation, 
        Bunsenstr. 10,
        D-37073 G\"ottingen, 
        Germany
  \\
  $^{2}$ Fachbereich Physik, 
         Philipps-Universit\"at Marburg,
         D-35032 Marburg, 
         Germany}

\ 

\begin{abstract}
  In shear flows like pipe flow and plane Couette flow there is an
  extended range of parameters where linearly stable laminar flow
  coexists with a transient turbulent dynamics.  When increasing the amplitude
  of a perturbation on top of the laminar flow, one notes a 
  a qualitative change in its lifetime, from smoothly varying and short 
  one on the laminar side to sensitively dependent on initial conditions and long
  on the turbulent side. The point of transition defines a point on the 
  edge of chaos. Since it is defined via the lifetimes, the edge of 
  chaos can also be used in
  situations when the turbulence is not persistent. It then
  generalises the concept of basin boundaries, which separate two coexisting attractors, 
  to cases where the dynamics on one side shows transient chaos and
  almost all trajectories eventually end up on the other side. In this paper we
  analyse a two-dimensional map which captures many of the features
  identified in laboratory experiments and direct numerical simulations of
  hydrodynamic flows.  The analysis of the map shows that different
  dynamical situations in the edge of chaos can be combined with
  different dynamical situations in the turbulent region.
  Consequently, the model can be used to develop and test further
  characterisations that are also applicable to realistic flows.
\end{abstract}


\submitto{New Journal of Physics}

\maketitle

\markboth{{Vollmer et al. --- Basin boundary, edge of chaos, and edge state in a 2d model}}{}

\section{Introduction}

The transition to turbulence in systems like plane Couette flow or
pipe flow differs from the better understood examples of
Taylor-Couette or Rayleigh-Benard flow in that turbulent dynamics is
observed while the laminar flow is still linearly stable
\cite{grossmann00,Ker05,Eck07b,Eckh08}.  Evidently, the two types of
dynamics coexist for the same parameter values. This suggests a subcritical
transition scenario, where the turbulent state forms around the node
in a saddle-node bifurcation. Indeed, various bifurcations of
saddle-node type have been found in these systems
\cite{Nag90,Nag97,Cle97,Wal03,WangGibsonWaleffe2007,Eck02,Fai03,TW_bristol,Pri07,aberdeen}
but at least in pipe flow they differ from the standard phenomenology
in that the node state is not stable but has unstable directions as
well: it is like a saddle-node bifurcation in an unstable subspace. 
Numerical studies of pipe flow \cite{Sch07a}and some
simplified models \cite{Sku06} show that also the `saddle state' has
peculiar features.  In the higher-dimensional space it need not be a
fixed point, as in the traditional saddle-node bifurcation scenario,
but can be dynamically more complicated, \ie periodic or even chaotic.

In Couette and pipe flow the turbulent state forming around the node
need not be an attractor.  Indeed, numerical and experimental evidence
indicates that at least in the transitional regime the turbulent dynamics
is not persistent but transient
\cite{brosa91,bead_bottin1,bead_bottin2,Moe04,FE04,Hof06,Mul05,Mul06,Pei06,Pei07,Schn08b}.
Nevertheless, it is still possible to define a boundary between
trajectories directly decaying into the laminar state and those first
visiting the neighbourhood of the chaotic saddle.
Trajectories on the turbulent side show a sensitive dependence on
initial conditions and give rise to rapidly varying lifetimes. This
suggested the name ``edge of chaos'' for this boundary \cite{Sku06}. 
In the case of the standard subcritical transition scenario, this
edge of chaos is given by the saddle state and its stable manifold
\cite{Ott2002}. There is some evidence that for such a behaviour in
plane Couette flow \cite{WangGibsonWaleffe2007,SchneiderGibson2007}. 
In the case of pipe flow numerical evidence
suggests that the saddle state is not a single fixed point or a
travelling wave, but that it rather carries a chaotic dynamics
\cite{Sch07a}. 

In order to explore some of the possibilities in a computationally
efficient and dynamically transparent manner, we turn to a specifically
designed model system. 
In the following we will describe a two-dimensional map that shows
much of the phenomenology observed in transitional pipe flow, and at
the same time has parameters that allow us to discuss the transitions
and crossover between different kinds of dynamical behaviour. We use
the model to study the boundary between laminar and turbulent
dynamics, and the dynamics in this boundary. In particular, we will
argue that the edge of chaos and the edge states introduced in
\citeasnoun{Sku06} and \citeasnoun{Sch07a} are the natural extension
of the basin boundary concept to situations where the turbulent dynamics
is transient.

Studying boundaries of basins of attraction has a long history in
dynamical systems. It goes back to Cayley for the case of Newton
iteration, and to Julia and Fatou for dynamical systems defined in the
plane of complex numbers \cite{peitgen,Devaney2003}.
To make contact with differential equations much follow up work
focussed on the conceptually simplest systems of flows in three
dimensions, or equivalently 2d invertible maps. In principle the
generic properties of the boundaries between the domains of attraction
of different types of invariant sets (sinks, saddles nodes, limit
cycles, and chaotic sets) have exhaustively been classified
\cite{RobertAlligoodOttYorke2000,Ott2002} for these systems by
considering (i) the possible sections of the respective stable and
unstable manifolds and (ii) the possible impact of
\hbox{(dis-)}appearance of stable orbits in saddle-node bifurcations.
However, careful inspections of the parameter dependence of
`explosions', where the features of invariant sets alter
qualitatively, can occasionally still unearth surprises in systems as
simple as the Hen\'on map \cite{Osinga2006}.
Higher-dimensional chaos (``hyperchaos'') shares common themes with low-dimensional
chaos \cite{Roessler1983}, but there also are important differences
due to the additional freedom of changing dynamical connections
between chaotic sets
\cite{GrebogiOttYorke1983PRL,LaiWinslow1995,DellnitzFieldGolubitskyHohmannMa1995,%
AshwinBuescuStewart1996,KapitaniakMaistrenkoGrebogi2003,RempelChianMacauRosa2004a,PazoMatias2005,TelLai2008}.
Besides fluid mechanics other important fields of applications of
hyperchaos are transition state theory
\cite{KovacsWiesenfeld2001,WigginsWiesenfeldJaffeUzer2001,WaalkensBurbanksWiggins2004,BenetBrochMerloSeligman2005}
and the quest for the (domain of) stability of irregular and stable
synchronised states in systems of coupled oscillators (see
\citename{PikovskyRosenblumKurths2001}
\citeyear{PikovskyRosenblumKurths2001} for an overview).  Considerable
insight in the latter problem come from studies of two symmetrically
coupled logistic maps
\cite{YamadaFujisaka1983,FujisakaYamada1983,GuTungYuanFengNarducci1984,%
PikovskyGrassberger1991,MaistrenkoMaistrenkoPopovichMosekilde1998,%
KapitaniakLaiGrebogi1999,KapitaniakMaistrenkoGrebogi2003}.
More recently also the generalisations to asymmetric coupling
\cite{HuYang2002,KimLimOttHunt2003} and more complex maps
\cite{Lai2001,KimLimOttHunt2003,AshwinRucklidgeSturman2004} have been
explored.

The present study is motivated by observations on the turbulence transition  
in situations where the laminar profile is linearly stable, and hence
will use descriptions like `laminar' and `turbulent' to describe the two
dominant state between which we would like to determine the basin boundary
or edge of chaos.
One of our principle interests will be in situations where the
dynamics on the edge of chaos separating (transient)
turbulence and laminar motion is chaotic. To that end our model must have at
least two continuous degrees of freedom --- one degree of freedom for
the dynamics in the edge, and a second one perpendicular to it. A
minimal model of the phase-space flow would then require at least 
a four-dimensional invertible map, but then we would loose the
advantages of the graphical representation of the invariant sets and their domains of
attraction that are available in lower dimensions.
As in the approaches to model synchronisation of coupled nonlinear
oscillators, we will therefore design a system of two coupled 1-d
maps.

The paper has three main parts. 
In the first part (\sect{2dmap}) we introduce the model, discuss the
dynamics of the uncoupled case, and introduce the considered coupling.
The second part (\sects{attractors} and \ref{sec:edgeState}) deals
with the dynamics of two coexisting attractors: In \sect{attractors}
we discuss the shape and dynamics of the attractors, and the transient
dynamics in the respective basins of attraction.  \Sect{edgeState}
addresses the dynamics of the relative attractor on the basin boundary
between the attractors, and how this dynamics effects the shape of the
separating boundary.
In the third part of the paper we turn to the case of a chaotic
repellor in the turbulent dynamics which mimics turbulent transients
decaying to a laminar flow profile: \Sect{transient} deals with the
case of a chaotic saddle coexisting with a fixed point attractor.  We
discuss the metamorphosis of the basin boundary at the crisis where
the attractor turns into a chaotic saddle.
Finally, in \sect{turbulence} we conclude the paper with summarising
remarks and discuss how the findings on this 2D model relate to
observations in shear flows such as turbulent pipe and
plane Couette flow.

\section{The two-dimensional map}
\label{sec:2dmap}

To admit coexistence of laminar and a turbulent dynamics one degree of
freedom of the map must be chosen along a phase-space direction
separating regions with these different types of dynamics.  A second
degree of freedom is needed to capture the dynamics perpendicular to
this direction, and to allow for dynamics \emph{within} the boundary
between laminar and turbulent dynamics.
We think of the two coordinates of the map as representing the energy
content of the perturbation ($x$-direction) and the dynamics in an
energy shell ($y$-coordinate).  The $x$-coordinate interpolates
between a laminar and a turbulent dynamics. The $y$-coordinate models
all other degrees of freedom. In the latter direction the map is
globally attracting towards a region near $y=1$. The combined map has
a fixed point, corresponding to the laminar profile, and --- for
suitable parameter values --- also a region with a chaotic dynamics
corresponding to turbulent behaviour.  In the following we first
describe the two uncoupled maps in $x$ and $y$, and then we discuss
their coupling and its consequences for the dynamics.

\subsection{Dynamics in $x$}

\begin{figure}
\rule{10mm}{0mm}
\includegraphics[width=0.45\textwidth]{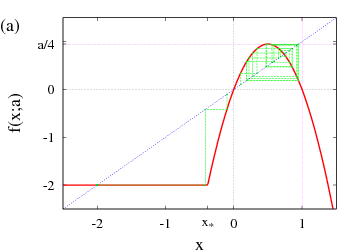}
\rule{10mm}{0mm}
\includegraphics[width=0.45\textwidth]{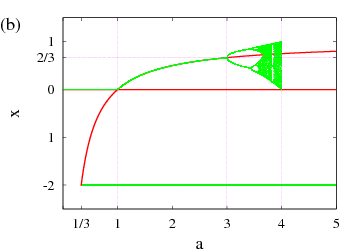}
\caption[]{\label{fig:x-map}
  The map along the $x$-direction.
  (a) The red line shows the function $f( x; \parX )$ for $\parX=3.8$,
  and the green ones indicate the evolution of two trajectories
  starting at $x=\pm 0.1$, respectively. 
  The maximum of the map is always at $x=1/2$, irrespective of the
  value of $\parX$, and takes the value $f(x=1/2; \parX) = a/4$.  
  The initial condition $0.1$ approaches the chaotic attractor in
  $[f(1/2);f^2(1/2)]=[0.18;0.95]$, and the one starting at $-0.1$
  approaches the fixed point at $x=-2$.
  (b)~Bifurcation diagram of the map  $f( x; \parX )$. Green dots represent points on 
  either of the two attractors of the map, and the red lines unstable fixed points. 
}
\end{figure}

For the dynamics along the energy axis, we use a map that has  
a stable fixed point at $x=-2$, and a chaotic dynamics for $x>0$.
The former corresponds to laminar flow, and the latter mimics turbulent
motion. An intermediate fixed point at $x=0$ separates the laminar
region $x<0$ from the turbulent one at $x>0$. It is unstable.  
These features are contained in the one-parameter map  [\Fig{x-map}(a)]
\numparts
\begin{equation}
x_{n+1}= f( x_n;\parX )
\end{equation}
with 
\begin{equation}
  f( x;\parX ) = 
  \left\{
    \begin{array}{ll}
      \parX \, x \, (1-x) \qquad & \textrm{for} \quad  
      x > x_* \equiv \left. \left( 1-\sqrt{1+8/\parX} \right)\right/2 \\
      -2 & \textrm{else}
    \end{array}
  \right. 
\label{eq:fxa}
\end{equation}
\endnumparts
Here $x_*$ is the leftmost intersection between the constant value
$-2$ for $x<x_*$, and the quadratic part at $x>x_*$. With this
choice the map is continuous.

The bifurcation diagram for this map is shown in \Fig{x-map}(b). 
We will only be interested in parameter values $\parX > 1/3$ where
$x_*>-2$. In this case the map has a stable fixed point at $x=-2$,
which absorbs all initial conditions starting outside the interval $[0,1]$.
Over the interval $x\in[0,1]$ the map coincides with the logistic map 
and shows its familiar bifurcation diagram.
For all $1/3 < \parX < 1$ there are stable fixed points at $x=-2$ and
$x=0$. In addition, there is an unstable fixed point at $x_s = 1-1/\parX$,
which lies between $-2$ and $0$.
At $\parX = 1$ the fixed point $x_s$ crosses $x=0$, and the two fixed
points change stability in a transcritical bifurcation. 
For $\parX > 1$ the point $x=0$ is unstable,
and $x_s$ is a stable fixed point.
At $\parX = 3$ the fixed point $x_s$ undergoes a first period doubling,
and subsequently follows the period-doubling route to chaos.
Beyond $a \simeq 3.59$ there are chaotic bands extending from 
$f(1/2;\parX)=a/4$ down towards
$f^2(1/2;\parX)=f(a/4;\parX)=(a/2)^2\,(4-a)$. 

At $\parX = 4$ the chaotic band generated by the period doubling
collides with the unstable fixed point at $x=0$, leading to a
\emph{boundary crisis}
\cite{greb82,GrebogiOttYorke1983,GrebogiOttRomeirasYorke1987,Ott2002}.
For $\parX>4$ some points near the maximum of the parabola are mapped
outside the interval $[0,1]$ and the attractor turns into a
chaotic saddle. All points except for a Cantor set of measure zero
will eventually map outside the interval and then be attracted to the
laminar fixed point at $x=-2$. The Cantor set contains an infinity of
orbits which follow a chaotic dynamics and never leave the interval
\citeaffixed{tel1990,Ott2002}{\cf}.

In summary, depending on the parameter values, the $x$-map shows the
coexistence of a stable laminar state with one of three possible types
of non-laminar dynamics: another fixed point, a chaotic attractor, or
a chaotic saddle.  The coexistence of a stable laminar fixed point at
$x=-2$ with a transient chaotic dynamics in the map for $\parX>4$
mimics the coexistence of a transient turbulent dynamics with a
linearly stable laminar steady flow.  The direct domain of attraction
of the laminar state at $x=-2$ is bounded towards positive $x$ by an
unstable fixed point at $x=0$.

\subsection{Dynamics in $y$}
\label{sec:y-dynamics}

\begin{figure}
\rule{10mm}{0mm}
\includegraphics[width=0.45\textwidth]{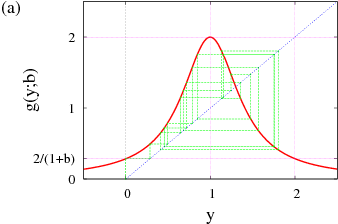}
\rule{10mm}{0mm}
\includegraphics[width=0.45\textwidth]{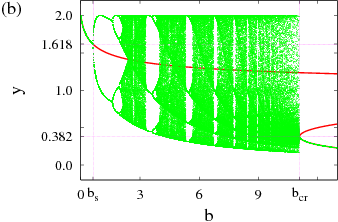}
\caption[]{\label{fig:y-map}
  The map in the $y$-direction.
  (a) The function $g ( y;\parY )$ for $\parY = 6$, and the
  trajectory of $y = 0$. 
  For all $\parY$ the maximum of $g ( y;\parY )$ is
  at $\left( y, g( 1;\parY ) \right)=(1;2)$.
  Moreover, $g ( 2;\parY ) = g ( 0;\parY ) = 2/(1+\parY)$
  such that the attractor of the map always lies in the interval 
  $[ 2/(1+\parY); 2 ]$.
  (b)~Bifurcation diagram of the map $g ( y;\parY )$. As in
  Fig.~\ref{fig:x-map} green dots represent points on the attractor,
  and the red lines mark two unstable fixed points of particular
  interest. }
\end{figure}

The $y$-dynamics represents the motion within the energy shell. In the
simplest case it is globally attracting towards a globally stable
fixed point. Then only the $x$-dynamics matters, and it represents the
dynamics along its unstable direction. In order to model the motion in
the energy shell we consider a unimodal (\ie a single-humped) map of
Lorentzian type (\Fig{y-map}(a)) that maps large $|y|$ towards the
region $y \simeq 1$,
\numparts
\begin{equation}
  y_{n+1}= g( y_n;\parY ) \, ,
\end{equation}
with
\begin{equation}
  g( y;\parY ) = \frac{2}{1 + \parY \, (y-1)^2 } \, .
\label{eq:gyb}
\end{equation}
\endnumparts
In its first iteration the map collects all initial conditions into
the interval $[0,2]$.  In this interval the map can have up to three
fixed points $y_p$.  For the discussion of the properties of the map
and the fixed points, it is convenient to solve the fixed point
equation for the parameter and to study
\begin{equation}
   \parY_0 (y_p) = \frac{2-y_p}{y_p \, (y_p-1)^2 } \, .
\label{eq:parY0}
\end{equation}
By evaluating $\rmd \parY_0 /\rmd y_p = 0$ one verifies that there is a
saddle-node bifurcation at the critical value 
$y_{cr} = \left( 3 - \sqrt{5} \right)/2 \simeq 0.382$. 
This corresponds to the parameter value
$ \parY_{cr} \equiv \parY_0 (y_{cr}) \simeq 11.09 $. 
Consequently, there is only a single fixed point for 
$\parY < \parY_{cr}$, and there are three fixed points 
for larger values of $\parY$. 

Making use of \Eq{parY0} in order to evaluate $\rmd g( y;\parY ) /\rmd y = -1$
one verifies that the single fixed point is stable for 
$ y > y_s = \left( 1 + \sqrt{5} \right)/2 \simeq 1.6182$,
\ie for 
$ \parY < \parY_s \equiv \parY_0 ( y_s ) \simeq 0.618 $.
Beyond $\parY_s$ the fixed point undergoes a period-doubling route into
chaos, and produces a broad chaotic band in the interval $[0,2]$. 
At $\parY_{cr}$ there is a saddle-node bifurcation in the support of
the attractor, which transforms the attractor into a saddle. For larger
values of $\parY$ this saddle coexists with a globally stable fixed point. 

\begin{figure}
\raisebox{0.3\textwidth}{(a)}
\includegraphics[height=0.35\textwidth]{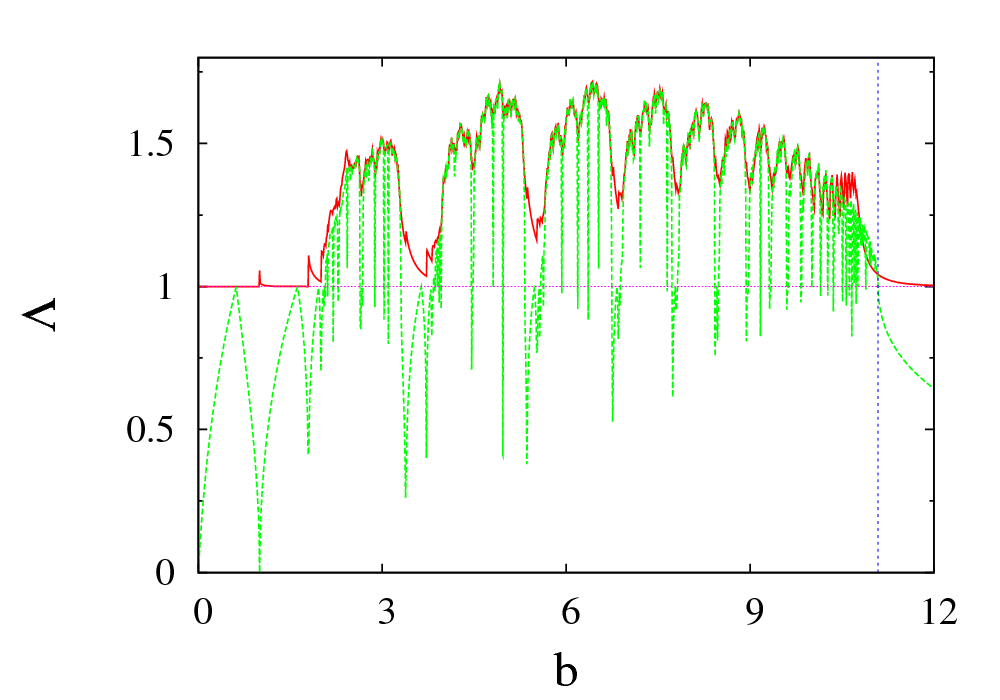}
\rule{6mm}{0mm}
\raisebox{0.3\textwidth}{(b)}
\rule{-2mm}{0mm}
\includegraphics[height=0.34\textwidth]{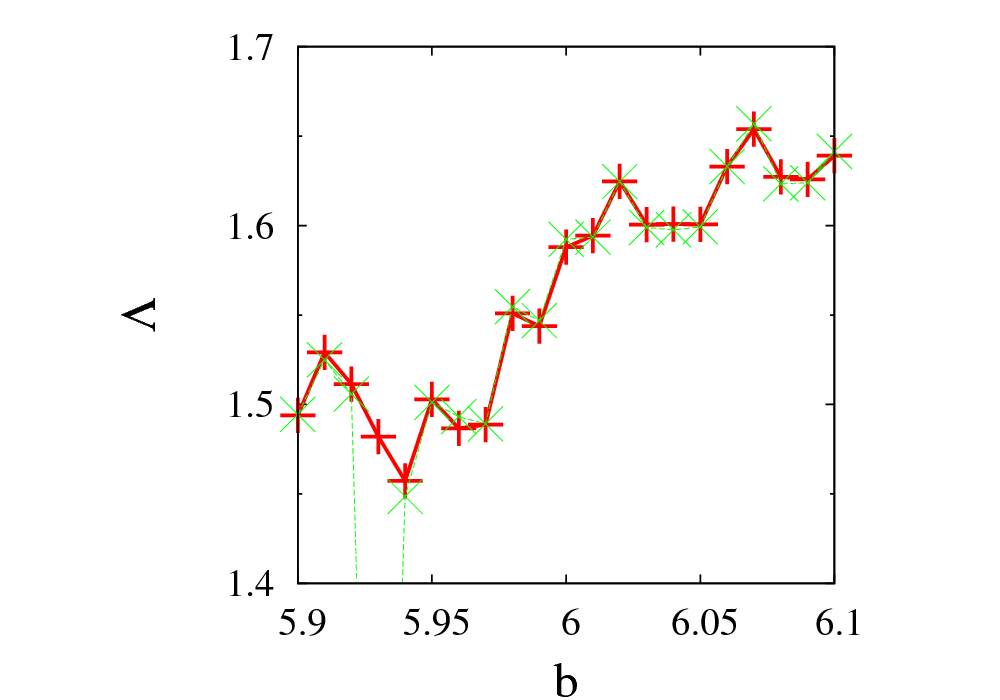}
\caption[]{\label{fig:thermodynamics}
  Lyapunov number $\Lambda$ of the attracting set (green) and the
  chaotic invariant set (red) of the map \eq{gyb}. The former number
  is obtained from the last $2\cdot 10^6$ iterates of a trajectories
  that is $10^7$ iterations long, and the latter by means of the
  thermodynamic formalism [\cf \citename{tel1988}
  \citeyear{tel1988,tel1990} for a description of the algorithm].  (b)
  Magnification of a small parameter interval to show that in regions
  where, within numerical uncertainty, there are no stable periodic
  orbits, the two Lyapunov numbers coincide.  }
\end{figure}

For later reference we introduce also the Lyapunov number $\Lambda$ of
the map, which describes how a small distance $\delta y_0 = | y^{(a)}_0 - y^{(b)}_0 |$
between two close-by initial conditions $y^{(a)}_0$ and $y^{(b)}_0$ grows with the
number $j$ of iterations,
\begin{equation}
\delta y_j 
\equiv 
| y^{(a)}_j - y^{(b)}_j | \sim \delta y_0 \; \Lambda^j \, .
\label{eq:Lambda}
\end{equation}
The Lyapunov numbers can be defined for invariant sets, such as the
maximal chaotic invariant set ($\Lambda_c$) and for the attractor
($\Lambda_a$). The distinction is important whenever the two numbers
do not coincide, as in cases where an attracting periodic orbit is
surrounded by an invariant chaotic set. 
The two Lyapunov numbers for the map \eq{gyb} are shown in
\Fig{thermodynamics}. The Lyapunov number for the maximal chaotic
invariant set is shown as a solid red line: it always remains above
$1$. 
The Lyapunov number of the attractor is shown by a dotted green
line. It takes values smaller than unity in the parameter windows
where there is an attracting periodic orbit. 

In summary, the main features of the $y$-dynamics are that it is globally
contracting towards the interval $[0;2]$, and that depending on the
parameter values one can have one of three types of invariant sets:
(i) a stable periodic orbit of period $2^n$ with $n=0$ (\ie a fixed
point) for $\parY < \parY_s$ and larger $n$ in the subsequent period
doubling cascade; 
(ii) a chaotic attractor for numerous parameters in the range 
$\parY_s < \parY < \parY_{cr}$; or 
(iii) a chaotic saddle coexisting with a periodic orbit (in the
periodic windows of the previous parameter regime) or a fixed point
for $\parY_{cr} < \parY$. 

\subsection{The coupling}

Without a coupling between the two maps, the three possibilities in
the $x$-dynamics combine with the three possibilities in the
$y$-dynamics for nine different regimes.  Now we introduce a coupling
between both degrees of freedom. The specific form of the coupling
should not be important if it preserves a few properties. For
instance, we want to keep a locally stable fixed point for the laminar
state also in the coupled dynamics.  Specifically, the $y$-map should
have a stable fixed point at $x=-2$.  We therefore introduce an
$x$-dependence in the parameter $\parY$ of the $y$-map such that the
coupling vanishes for $x \simeq -2$, thereby maintaining the stability
properties of the uncoupled map:
\begin{equation}
  \parY(x) = \gamma \, (2+x) \, .
\label{eq:parYx}
\end{equation}
We refer to this fixed point as the \emph{laminar fixed point}.
Since the non-trivial $x$-dynamics lies within the interval $[0,1]$,
the range of $\parY$ values varies between $2\gamma$ and $3\gamma$, so
that the parameter $\gamma$ selects the type of $y-$dynamics for the
chaotic regime in the $x$ dynamics.

To complete the coupling we also introduce an influence of the
$y$-dynamics on the $x$-dynamics, since otherwise 
the bifurcations are
determined by the $x$-map alone: we shift $x_n$ by the deviation of
$y_n$ from the position of the maximum before applying the mapping,
\ie 
\numparts
\begin{eqnarray}
  x_{n+1}&=& f(x_n  - \epsilon \, ( y_n - 1 ); \parX) 
  \label{eq:coupled_x}
  \\
  y_{n+1}&=& g( y_n; \parY(x_n) )
  \label{eq:coupled_y}
\end{eqnarray}
\label{eq:coupled_evo}
\endnumparts
with the specific forms (\ref{eq:fxa}), (\ref{eq:gyb}), and (\ref{parYx})
for $f(x)$, $g(x)$, and $b(x)$, respectively.
In this paper we will concentrate on the weak-coupling limit where
$\epsilon \ll 1$. Unless stated otherwise this parameter will always
take the value $\epsilon=0.03$.

This completes our definition of the coupled map.
Through appropriate choices of the parameters $\parX$ and $\gamma$ we
can --- one by one --- study the nine parameter regimes with their
qualitatively different dynamics.  We here begin with the six cases
where the non-laminar $x$-dynamics is attracting, and a laminar and a
non-laminar attractor coexist.  The case of a transient dynamics will
be taken up in \sect{transient}.

\section{Two coexisting attractors}
\label{sec:attractors}

\begin{figure}
\rule{10mm}{0mm}
\raisebox{42mm}{(a)}
 \includegraphics[width=0.4\textwidth]{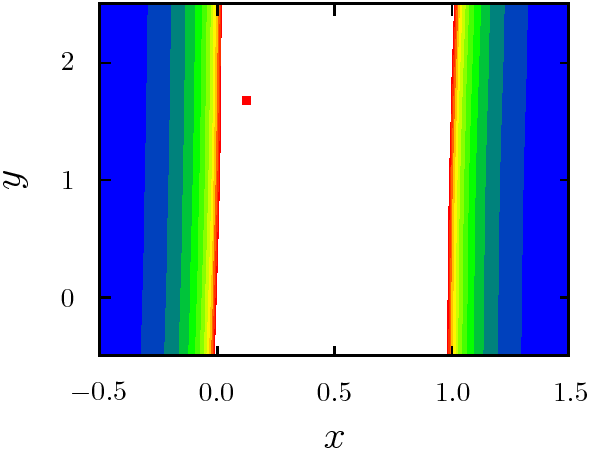}
\rule{10mm}{0mm}
\raisebox{42mm}{(d)}
 \includegraphics[width=0.4\textwidth]{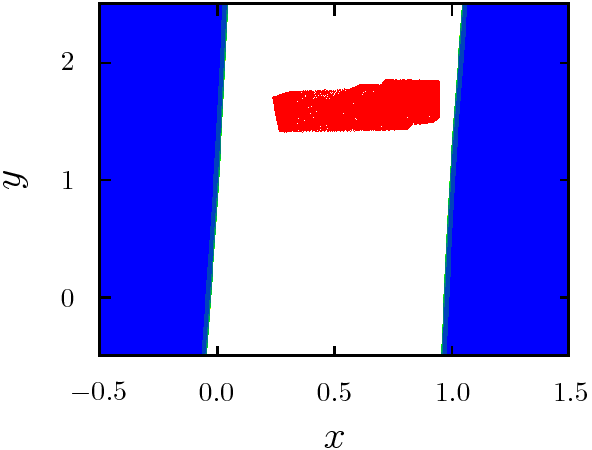}
\\[8mm]
\rule{10mm}{0mm}
\raisebox{42mm}{(b)}
 \includegraphics[width=0.4\textwidth]{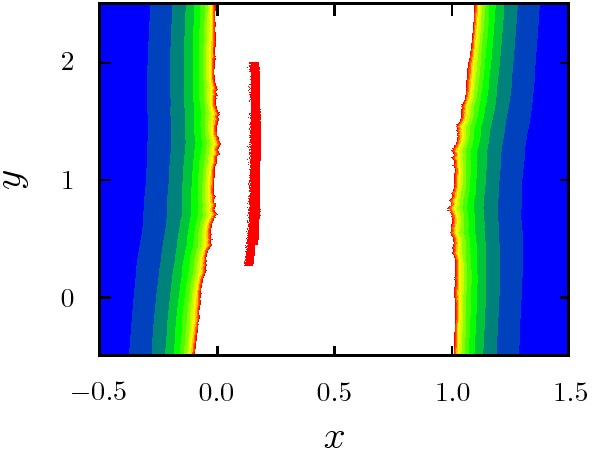}
\rule{10mm}{0mm}
\raisebox{42mm}{(e)}
 \includegraphics[width=0.4\textwidth]{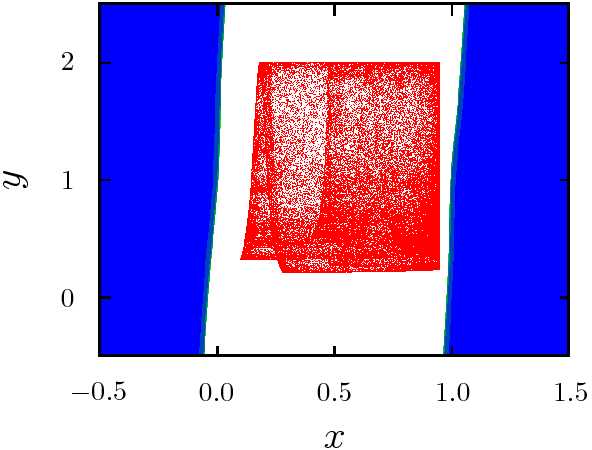}
\\[8mm]
\rule{10mm}{0mm}
\raisebox{42mm}{(c)}
 \includegraphics[width=0.4\textwidth]{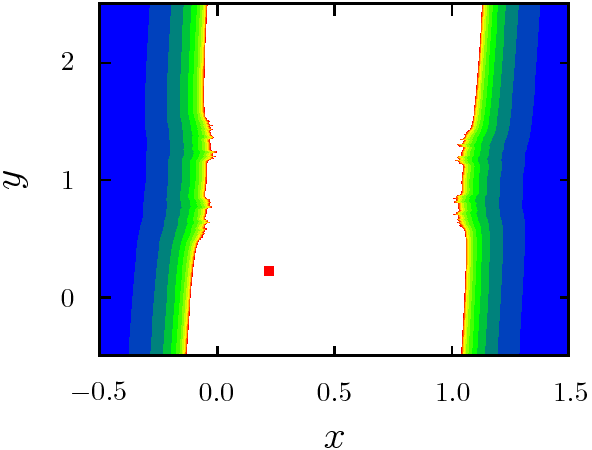}
\rule{10mm}{0mm}
\raisebox{42mm}{(f)}
 \includegraphics[width=0.4\textwidth]{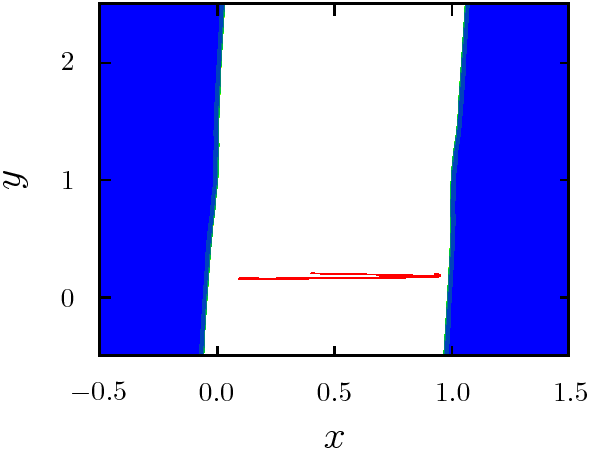}
\caption[]{\label{fig:invSets} For initial conditions $(x,y)$ the
    colours indicate the number of iterations required
    to reach the laminar fixed point. The colour code runs from one
    iteration (blue) to more than ten iterations (red). 
  Initial conditions in the white region are attracted to the chaotic
  attractor, which is also shown by red dots. 
  The panels in different rows refer to different values of $\gamma$:
   (a,d) $\gamma=0.2$, (b,e) $\gamma=3$, and (c,f) $\gamma=6$,
   respectively.  In these cases the $y$-dynamics shows a single fixed
   point, chaos, and a fixed point coexisting with a chaotic saddle,
   respectively.
   The left panels (a--c) and right panels (d--f) refer to $\parX =
   1.2$ and $\parX=3.8$, respectively.  
   For the panels (a--c) the non-laminar $x$-dynamics
   amounts to a fixed point, and for (d--f) it is chaotic. 
  In all panels $\epsilon$ is set to $0.03$, except for the top left
  one, where $\epsilon=0.01$, because in the latter case the
  non-trivial fixed point at $x>0$ disappears for $\epsilon \gtrsim 0.012$.
}
\end{figure}
\subsection{Attractors and basins}

Figure~\fig{invSets} shows the domain of attraction of the laminar
fixed point together with the non-laminar attractor.
The panels on the left-hand side refer to $\parX=1.2$ immediately beyond
the crossing of stability, where $f(x,\parX)$ has a stable fixed-point
attractor at $x=1-1/1.2\simeq 0.17$.
The panels on the right-hand side refer to $\parX=3.8$. In this case
the fixed point at $x=-2$ coexists with a chaotic attractor. Without
coupling, for \eq{coupled_x} with $\epsilon=0$ the attractor is
located in the interval $[0.18;0.95]$.

For $\gamma = 0.2$ [\Fig{invSets}(a,d)] the iteration of $y$ directly
approaches the fixed point at $y \simeq 1.8$, and subsequently only
wiggles around this point due to the perturbation arising from the
$x$-dynamics. 

For $\gamma = 6$ [\Fig{invSets}(c,f)] the iteration of $y$
approaches the fixed point at $y \simeq 0.38$. However, in this case
the fixed point is surrounded by a chaotic saddle, and the
approach may involve long chaotic transients.

Finally, for $\gamma = 3$ [\Figs{invSets}(b,e)] the parameter $\parY(x)$
varies between $6$ and $9$ for $x$-values in the interval $[0,1]$. 
For parameter values $\parX$ slightly below $4$ one hence obtains a
chaotic dynamics for both $x$ and $y$. In this case the
attractor varies over a considerable range of $y$-coordinates, and one
can clearly see the strong influence of the coupling. When there are
broad chaotic bands in both directions [\Fig{invSets}(e)] the attractor
can even extend to negative values of $x$. 

\begin{figure}
\rule{10mm}{0mm}
\raisebox{42mm}{(a)}
 \includegraphics[width=0.4\textwidth]{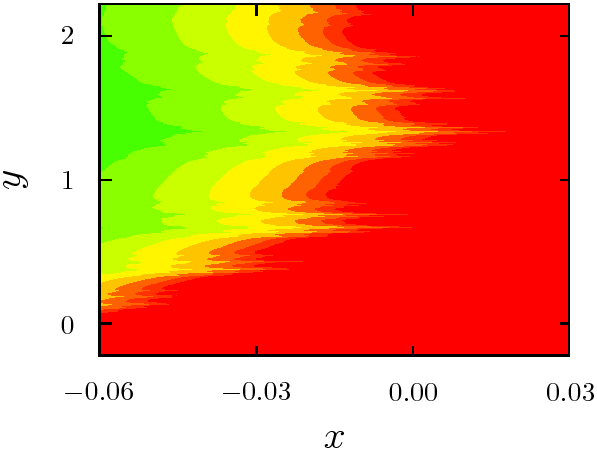}
\rule{10mm}{0mm}
\raisebox{42mm}{(b)}
 \includegraphics[width=0.4\textwidth]{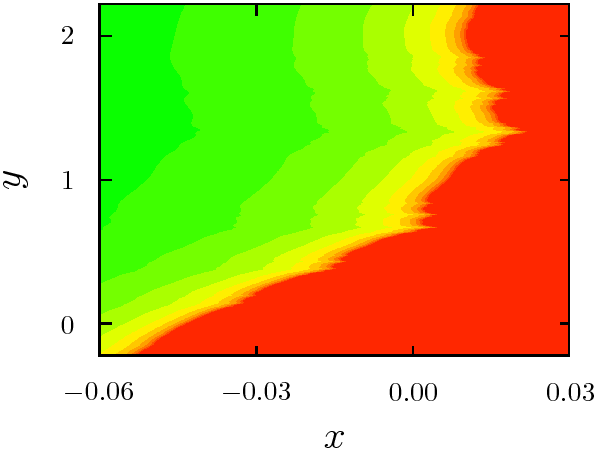}
\\[8mm]
\rule{10mm}{0mm}
\raisebox{42mm}{(c)}
 \includegraphics[width=0.4\textwidth]{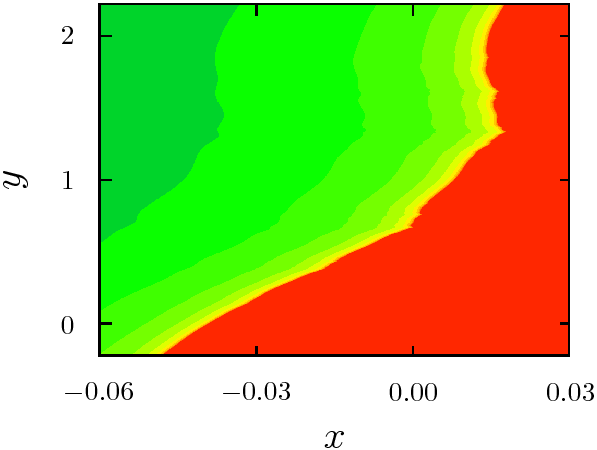}
\rule{10mm}{0mm}
\raisebox{42mm}{(d)}
 \includegraphics[width=0.4\textwidth]{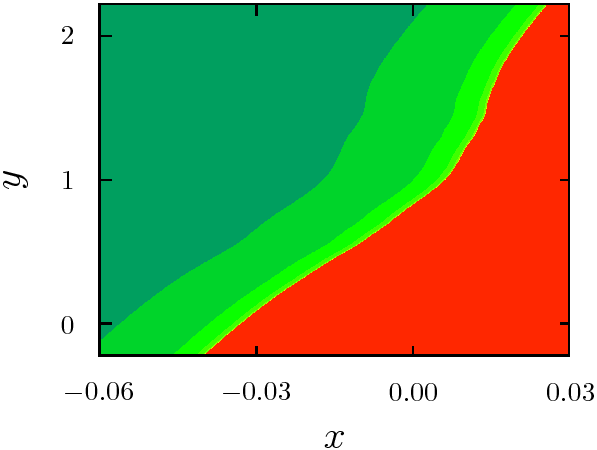}
\caption[]{\label{fig:rough_border}
  Magnifications of the boundary between the region of initial conditions
  approaching the laminar fixed point at $(-2,2)$, 
  and the chaotic attractor at $x>0$
  for $\gamma = 3$, $\epsilon=0.03$, and different values of $\parX$: 
  (a)~$\parX=1.2$, (b)~$\parX=1.6$, (c)~$\parX=2.0$, (d)~$\parX=3.8$. 
  All initial conditions which approach the attractor to the right
  are shaded in red, and the colour gradient from blue to yellow
  indicates the number of steps it takes to reach the laminar fixed
  point with the colour coding also used in \Fig{invSets}.
  The cross-over from a rough to a
  smooth boundary arises at $\parX \simeq 1.6$.  }
\end{figure}
\subsection{The boundary between the two attractors}
\label{sec:fractal_boundary}

We now focus on the boundary separating the basins of attraction of
the laminar and the chaotic attractor to the left and right,
respectively.  If $\epsilon=\gamma=0$, it coincides with the $y$-axis:
All initial conditions with $x>0$ are attracted to the turbulent
dynamics, and the ones with $x<0$ to the laminar state. Moreover, all
points with $x=0$ are immediately mapped into the hyperbolic fixed
point $(0,2)$.  The hyperbolic fixed point then becomes a relative
attractor, since it is an attractor for initial conditions in the
boundary between the two attractors.  For $\epsilon$ and $\gamma$
nonzero but small, the hyperbolic point is slightly shifted, and the
boundary no longer coincides with the $y$-axis, but it remains smooth.
The boundary can be determined by picking initial conditions with,
say, prescribed $y$ and varying $x$ and following them for some
iterations forward in time:
it can then be bracketed by a pair of $x$-values where one initial
condition iterates towards the laminar state and the other towards the
turbulent one.  This method allows us to track the dynamics in the
boundary not only in the case where the relative attractor is a fixed
point, but also when it is more complicated.
In a hydrodynamic setting this approach has been explored in the
framework of low-dimensional shear flow models \cite{Sku06} and direct
numerical simulation of pipe flow \cite{Sch07a}.
 
In \Fig{invSets} the boundary between the two attractors is the
boundary of the region shaded from blue to yellow. It appears to be
smooth for $\parX=3.8$ and for $\gamma=0.2$. In contrast, it looks
irregular for $\parX=1.2$ and $\gamma=3$ or $6$ [\Figs{invSets}(b,c)].
The magnifications in \Fig{rough_border} confirm the roughness of the
boundary and indicate a crossover from a smooth to an irregular
boundary as $\parX$ decreases from $3.8$ to $1.2$, with $\gamma=3$ and
$\epsilon=0.03$ fixed.

There are two elements needed to understand the emerging roughness of
the boundary: the first one is the observation that states in the
boundary are attracted to a subset of the boundary itself, \ie the
dynamics in the basin boundary converges to an \emph{edge state}.  The
second observation is that when the \emph{edge state} is chaotic a
rough boundary can form provided that the Lyapunov exponent for the
chaotic motion on the basin boundary is larger than the one
characterising the escape from the boundary. These two aspects are
discussed in the next section.

\section{Edge states and relative attractors}
\label{sec:edgeState}

\subsection{Identifying the edge state}

In order to follow a trajectory for long times and to be able to
identify the relative attractor, the bracketing of trajectories in the
boundary described in \sect{fractal_boundary} has to be refined after
some time. After all, the distance between the trajectories in the
pair bracketing the trajectory on the boundary grows exponentially
with the number of iterations.  Specifically, we proceed as follows.
We take initial conditions for the two trajectories that have equal
$y$-values and $x$-values separated by less than $10^{-6}$. The two
trajectories are followed until $\max(\delta x_j,\delta y_j)$ exceeds
$5\cdot 10^{-3}$.  Then a new pair is determined with $y_0 = (y^{a}_j
+ y^{b}_j)/2$ and $\delta x_0 < 10^{-6}$.
An alternative approach could start from the observation that the line
connecting the two trajectories will be oriented along the direction
of the largest Lyapunov exponent of the map and search for a
refinement along this line. Here and in the previous applications to
pipe flow \cite{Sch07a,SchneiderEckhardt2008} it was observed that the first approach, which
repeatedly projects the line segment between the two points to a fixed
direction in space, is more robust and converges more reliably,
especially in cases where the geometry of the boundary is complex.

\begin{figure}
\rule{25mm}{0mm}
\raisebox{0.20\textwidth}{(a)}
 \includegraphics[width=0.33\textwidth]{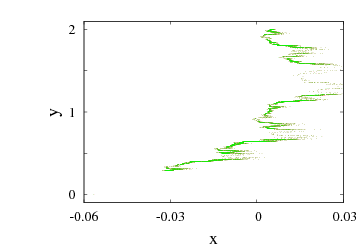}
\rule{10mm}{0mm}
 \includegraphics[width=0.33\textwidth]{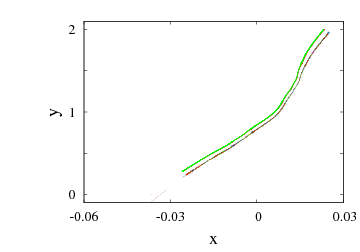}
\\[4mm]
\rule{25mm}{0mm}
\raisebox{0.20\textwidth}{(b)}
 \includegraphics[width=0.33\textwidth]{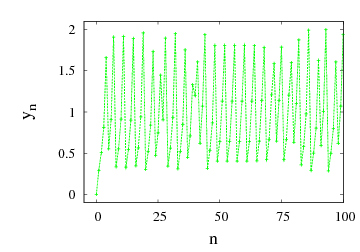}
\rule{10mm}{0mm}
 \includegraphics[width=0.33\textwidth]{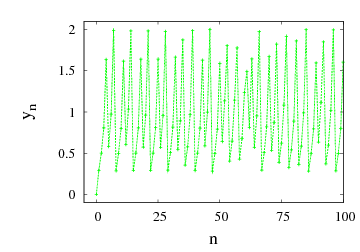}
\\[4mm]
\rule{25mm}{0mm}
\raisebox{0.20\textwidth}{(c)}
 \includegraphics[width=0.33\textwidth]{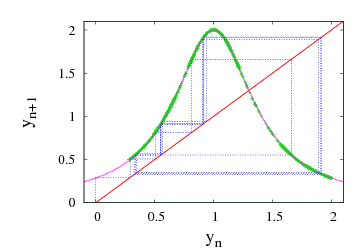}
\rule{10mm}{0mm}
 \includegraphics[width=0.33\textwidth]{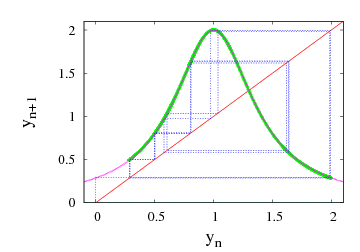}
\\[4mm]
\rule{25mm}{0mm}
\raisebox{0.20\textwidth}{(d)}
 \includegraphics[width=0.33\textwidth]{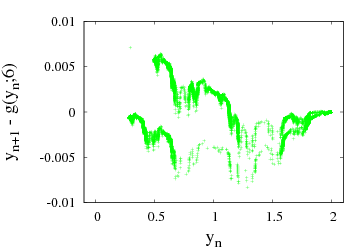}
\rule{10mm}{0mm}
 \includegraphics[width=0.33\textwidth]{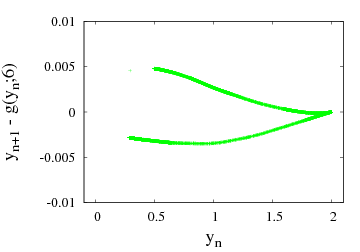}
\\[4mm]
\rule{25mm}{0mm}
\raisebox{0.20\textwidth}{(e)}
 \includegraphics[width=0.33\textwidth]{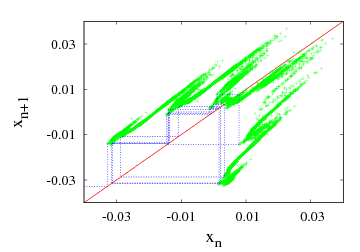}
\rule{10mm}{0mm}
 \includegraphics[width=0.33\textwidth]{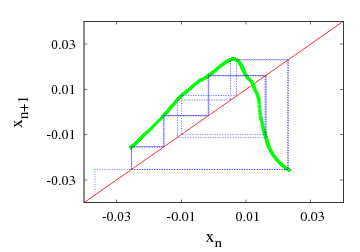}
\caption[]{\label{fig:rel_attractor}
Evolution of trajectories on the boundary separating convergence to
laminar and to turbulent motion ($\gamma=3$ and $\epsilon=0.03$ for
all panels; left panels: $\parX = 1.2$; right panels: $\parX = 3.8$).
(a) The trajectories on the boundary trace out the shape of the boundary,
which is rough for $\parX=1.2$ and smooth for $\parX=3.8$. 
For later reference the right panel also shows the edge of chaos for
$\parX=4.2$ beyond the crisis of the attractor. It has been shifted by 
$(x;y) = (0.001; -0.04)$ to the lower right. 
(b) When started at $x=0$ the trajectories rapidly converge to a
chaotic attractor located in the interval $y \in [2/7;\,2]$.
(c) Return map for the $y$-coordinate.
Within numerical accuracy it agrees with $g(y; 6)$ (solid
red line). The first ten iterations are explicitly indicated --- the
points visited during the initial 10000 time steps are indicated by
green crosses ($+$). 
(d) Deviation of the values plotted in (c) from the function  $g(y;6)$.
(e) Return map for the $x$ coordinate. The first iterations and
subsequent points indicated as in panel (c).  }
\end{figure}

The dynamics in the edge state is explored further in
\Fig{rel_attractor}. Presented are two situations where the boundary
shown in \Fig{rough_border} appears smooth ($a=1.2$, left column) and
rough ($a=3.8$, right column), respectively. The two frames in
\Fig{rel_attractor}(a) show trajectories on the boundary constructed
by the edge tracking algorithm. The trajectories nicely reproduce the features
of the boundaries also shown in \Fig{rough_border}(a) and (d). 
The difference between the two figures is that the boundary in
\Fig{rough_border} emerges from a two-dimensional search, whereas the
one in \fig{rel_attractor} is determined by a following a single
trajectory. This allows us to show
the time series of the coordinates of 
edge trajectories and the associated return maps for the
$y$-coordinates 
in row (b) and (c), respectively.  By visual inspection it is
very hard to see differences 
to the unperturbed dynamics of $g(y;b(0))$.

To demonstrate 
effects introduced by the coupling of the dynamics 
to the unstable $x$-direction we subtract the functional form of the $y$-map. 
The deviations from the unperturbed 
$y$-dynamics, $\delta y_n=y_{n+1}-g(y_n)$, 
differ substantially for smooth and rough boundaries:
For $\parX=3.8$ the iterates lie on a smooth, double valued curve.
Its double-valuedness reflects the influence of a non-trivial
dynamics in $x$, which follows iterates of a map with a single
bump, see the iterates in row (e). However, the relation between 
$x$ and $y$ is single valued, and therefore there is not much disorder.
For $\parX=1.2$, the distribution of iterates looks rather noisy (e),
and no simple relation between their images can be found. 

Note that in both cases the dynamics in $y$ is chaotic, and along the
$x$-direction close-by trajectories escape exponentially from the
vicinity of the boundary: both Lyapunov numbers are positive.  On the
other hand, for $\parX=3.8$, the different branches of the return map
come to lie on a smooth invariant set, while for smaller $\parX$ the
basin boundary is a rough invariant set.  In the next subsection we
argue that this difference is due to a crossover of the absolute
values of the Lyapunov numbers, just as it has been discussed in the
context of unstable-unstable pair bifurcations
\cite{GrebogiOttYorke1983PRL,TelLai2008}.

\subsection{Transition between smooth and rough boundaries}
\label{sec:roughTransition}

Close to $\parX=1.6$ the boundary crosses over from a highly irregular
geometry to a line with only few kinks, whose number progressively
decreases for even larger values of $\parX$. Similar transitions
between rough and smooth boundaries have previously been seen in
unsteady-unsteady pair bifurcations
\cite{GrebogiOttYorke1983PRL,Ott2002} and phase-synchronised chaos
\cite{Hunt97,Rosa1999}. They are related to a crossover of the two
Lyapunov numbers of the map. 
To gain insight into the transition we estimate the slope %
$| \delta y_0 / \delta x_0 |$ 
of the boundary at a point $(x_0,y_0)$.
Linearising \Eq{coupled_x} around the iterates $(x_j, y_j)$ of the
considered point we find 
\begin{equation}
\delta x_{j+1} 
\equiv  f'\left( x_j - \epsilon ( y_j - 1) \right) 
\; \left( \delta x_j - \epsilon \delta y_j \right)
= \sigma_{j} \; \left( \delta x_j - \epsilon \delta y_j \right)
\label{eq:DeltaXj}
\end{equation}
where $\sigma = f'(x) = a\,(1-2x)$ is the derivate of $f(x;a)$.
Since all points lie on the boundary, $x_j - \epsilon \, ( y_j - 1)$
is close to zero for all $j$, as can also be verified by inspection of
\Figs{invSets} and \fig{rough_border}.  Consequently, $f'$ is always
evaluated at a point close to zero, and $\sigma_{j}$ takes values
close to $a$.
By recursively working out \Eq{DeltaXj} we find
\begin{eqnarray}
& \delta x_n = &
\left( \prod_{k=1}^n \sigma_{n-k} \right) \; \delta x_0
-
\epsilon \; \sum_{j=1}^n \; 
\left( \prod_{k=1}^j \sigma_{n-k} \right) \;
\delta y_{n-j}
\nonumber\\[4mm]
\Longleftrightarrow \quad 
& \delta x_0 = & 
\frac{ \delta x_n }%
{ \prod_{k=1}^n \sigma_{n-k} }
+
\epsilon \; 
\sum_{j=0}^{n-1} \;
\frac{ \prod_{k=1}^j \sigma_{n-k} }%
{ \prod_{k=1}^n \sigma_{n-k} }
\;
\delta y_{n-j}
\end{eqnarray}

For an initial perturbation which is located on the boundary the
deviation $\delta x_n$ is bounded, and --- in the present case --- in
absolute value it is much smaller that unity.
(\cf\Fig{rough_border}).  On the other hand, for large $n$ the
denominator $\prod_{k=1}^n \sigma_{n-k}$ takes on very large
values --- after all, $a > 1$ and $\sigma_{k} \simeq 1$. Consequently,
\begin{equation}
\delta x_0 \simeq 
\epsilon \; 
\sum_{j=0}^{n-j-1} \;
\frac{ \delta y_{n-j} }%
{ \prod_{k=0}^{j-1} \sigma_{k} }
\label{eq:delta_x0}
\end{equation}

In the limit of very small perturbations and large $n$ we can
approximate the product by its asymptotic scaling, \ie
\[
\prod_{k=0}^{n-1} \sigma_{k} \sim a^n \, .
\]
In addition, according to \Eq{parYx} the parameter $\parY$ of $g$
always takes values very close to $2\gamma$ because all $x_j$ are very
close to zero. As shown in \Fig{rel_attractor}(c) the dynamics of the
$y$ coordinate essentially amounts to the unperturbed dynamics such
that we may use \Eq{Lambda} to related $\delta y_j$ to $\delta y_0$. 

In the scaling regime the sum in \Eq{delta_x0} can be worked out,
yielding 
\begin{equation}
\left |  \frac{ \delta x_0 }{ \delta y_0 } \right |
\sim 
\epsilon \: H \; \frac{ H^n - 1 }{H - 1} 
\qquad \hbox{ with } \quad 
H = \frac{ \Lambda }{a} \, .
\label{eq:slope}
\end{equation}
In the limit $n\rightarrow \infty$ the right hand side of \Eq{slope}
remains finite only if $H < 1$.  Hence, the boundary will be
smooth for $H<1$, or $\Lambda < a$.  On the other hand the
bound~\eq{slope} diverges for $H>1$. In this case the slope diverges
at least for some points on the boundary, which will hence be rough.%
\footnote{ A discussion of the abundance and distribution of singular
  points, and the fractal dimension of the basin boundary lies beyond
  the scope of the present manuscript. They can explicitly be worked
  out along the lines indicated in \citeasnoun{Rosa1999}.}

As noted above for points on the boundary the parameter $\parY$ of $g$
always takes values very close to $2\gamma$. According to
\Fig{thermodynamics} one thus finds that $\Lambda \simeq 1.59$ for
$\gamma=3$. The crossover from a rough to a smooth boundary should
therefore occur at $a \simeq 1.59$, which is in excellent agreement
with the numerical findings of \Fig{rough_border}.

This completes the characterisation of the attractors and their basin
boundary. In the following section we address the case of a chaotic
saddle coexisting with a laminar fixed point.

\section{Transient chaos}
\label{sec:transient}

\subsection{Lifetime Plots}

\begin{figure}
\rule{-10mm}{0mm}
\hbox{
\raisebox{31mm}{(a)}
\rule{-3mm}{0mm}
\includegraphics[width=0.3\textwidth]{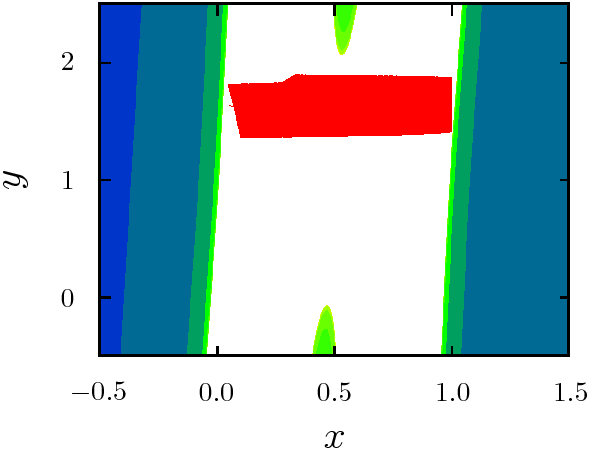}
\rule{4mm}{0mm}
\raisebox{31mm}{(b)}
\rule{-3mm}{0mm}
\includegraphics[width=0.3\textwidth]{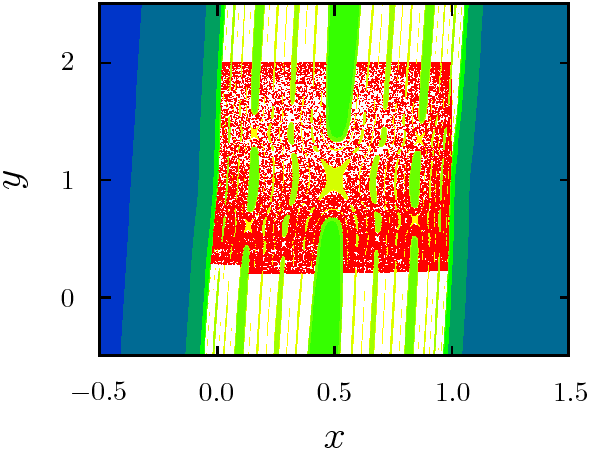}
\rule{4mm}{0mm}
\raisebox{31mm}{(c)}
\rule{-3mm}{0mm}
\includegraphics[width=0.3\textwidth]{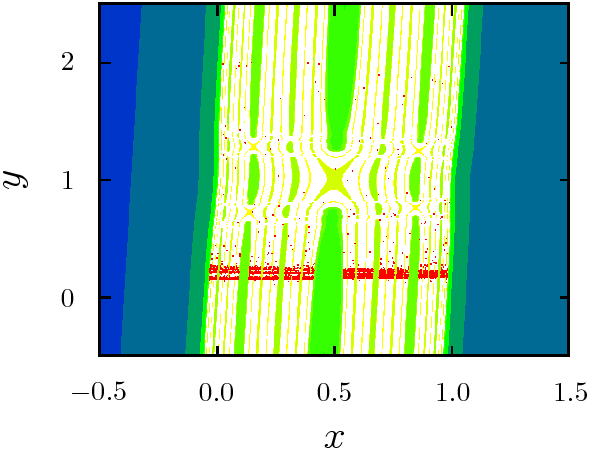}
}
\caption[]{\label{fig:saddle}
  The change of the structure of the invariant chaotic set (red dots)
  when the attractor undergoes crisis. 
  The colour coding indicates initial conditions arriving at the
  laminar fixed point in at most $10$ iterations. The tenth
  iteration of points which have not yet reached the laminar fixed
  point after $40$ iterations are indicated by red boxes. 
  Parameter values are $\parX=4.0$ and $\epsilon=0.03$ for all panels,
  while $\gamma$ takes on different values: (a) $\gamma=0.2$
  immediately before the transition from the attractor to the saddle,
  (b) $\gamma=3$, and (c) $\gamma=6$.  }
\end{figure}

The six cases discussed in the preceding sections cover the cases of
coexisting attractors. However, close to the transition in plane
Couette flow and pipe flow the turbulent dynamics is transient, so
that also the cases of a coexistence between a laminar fixed point and
a chaotic saddle that supports transient chaotic dynamics are of
interest. Our map realizes this for $\epsilon=0.03$ and $a \gtrsim 4$
(see \Fig{saddle}).  As in \Fig{invSets} we consider the three cases
(a) $\gamma=0.2$, (b) $\gamma=3$ and (c) $\gamma=6$.

When the parameter $\parX$ exceeds a critical value
$\parX_{cr}(\gamma)$, the laminar fixed point becomes globally
attracting except for a measure zero set containing periodic and
aperiodic trapped orbits left over from the attractor. This is
apparent in the plots in \Fig{saddle}, which show the lifetime of
initial conditions $(x,y)$ for $\parX=4.0$ and different values of
$\gamma$.  For $\gamma=0.2$ [\Fig{saddle}(a)] the critical value
$\parX_{cr}$ is larger than $4.0$, \ie there still is a stable chaotic
attractor coexisting with the laminar fixed point. However, we already
see two `fingers' approaching the attractor from the top and from the
bottom.
When increasing either $\gamma$ or $\parX$ these fingers are joined by
additional narrower fingers which all simultaneously collide with the
attractor at the parameter value $\parX_{cr}(\gamma)$.  Beyond this
\emph{crisis} most of the trajectories of the former attractor escape
through the regions where the collision took place \cite{Ott2002}.
The orbits of the attractor which never enter the regions form a
chaotic saddle.

The panels \Fig{saddle}(b,c) show the situation beyond the crisis. 
The blue areas iterate to the laminar fixed point in one and two
iterates, respectively.
The dark green strips near $x \simeq 0$ and $x \simeq 1$ arrive at
the fixed point in three iterations, and
initial conditions in the widest fingers (also dark green) pointing towards
$(x,y)=(0.5;1)$ escape to the laminar fixed point in four iterations. 
On the next level there are four lighter green fingers lying between
the widest fingers and the outer regions ($0 < x < 0.5$ and $0.5 < x <
1$), respectively, which are mapped to the fingers near $x \simeq 0.5$.
With each additional iteration, the number of fingers doubles.  
At the crisis all fingers simultaneously collide with points lying at the
upper and lower boundaries of the attractor. They can be interpreted as a
primary collision of the attractor with its basin boundary, and the
simultaneous collision of all the pre-images of this point.

What happens to the basin boundary of the attractor when going through
the crisis? The chaotic attractor embedded in the basin boundary
merges with the attractor. We have seen that this generates a fractal
set of ``holes'' (actually the fingers) through which trajectories of
the former attractor escape to the laminar fixed point. The chaotic
trajectories that never enter the fingers form a Cantor set. Since
trajectories starting in the domain of attraction are attracted
towards (a small neighbourhood of) the Cantor set and those starting
in the vicinity of this set escape almost certainly to the laminar
state, the Cantor set forms a chaotic saddle for the dynamics.  There
are orbits approaching this set from outside, but randomly selected
points in the vicinity of every point of the Cantor set eventually
approach the laminar state with probability one.

Figure~\ref{fig:rel_attractor}(a) shows orbits on the boundary
separating the respective domains of attraction towards the laminar
fixed point and the chaotic set.  As demonstrated in
\Fig{rel_attractor}(a, right panel) these orbits change smoothly
when the system undergoes crisis. The transition from a system with a
chaotic attractor to one with only chaotic transients is solely
reflected in the fact that the orbits on the edge of chaos attain new
pre-images. Their forward dynamics is not affected. In this respect
the trajectories forming the basin boundary remain a well-defined set
also beyond crisis. Their closure is the \emph{edge of chaos.}

Most initial conditions from the former attractor sooner or later
cross the edge of chaos. On the other hand the close-by points on the
Cantor set, which forms the chaotic saddle, never cross the edge of
chaos.  Some of them step on the edge and are attracted towards the
relative attractor on the edge of chaos. They give rise to the
additional pre-images mentioned above. Most points of the Cantor set,
however, only closely approach the edge of chaos, and subsequently
follow its unstable directions to explore the full support of the
Cantor set.
In this sense the edge of chaos remains a well-defined object also
after the crisis. It separates initial conditions where \emph{all}
orbits immediately decay to the laminar fixed point from a region
where they can perform a chaotic transient --- either short but
occasionally also very long.  In this sense the edge of chaos
separates initial conditions which are characterised by their
different finite-time dynamics rather than by their asymptotic behaviour: the
notion of the edge of chaos extends the concept of a basin boundaries
between two attractors to the situation of an attractor coexisting
with a chaotic saddle.

\subsection{Parameter dependence of the lifetime for initial conditions on the $y$ axis}

A useful and experimentally accessible indicator for the boundaries
and their dynamics are lifetimes of perturbations. \Fig{saddle} shows
the lifetimes for fixed parameters and a two-dimensional domain of
different initial conditions. The frequently used lifetime plots for
turbulence transitions differ from this one in that they usually show
the deviations for a combination of one coordinate (the amplitude of a
velocity field) and a parameter (the Reynolds number).

\begin{figure}
\rule{-10mm}{0mm}
\hbox{
\raisebox{31mm}{(a)}
\rule{-3mm}{0mm}
\includegraphics[width=0.3\textwidth]{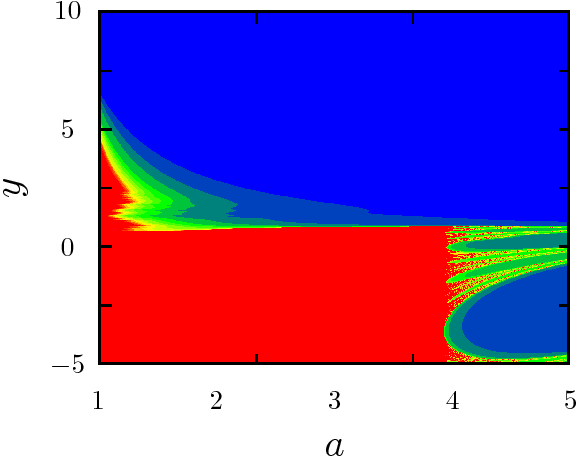}
\rule{4mm}{0mm}
\raisebox{31mm}{(b)}
\rule{-3mm}{0mm}
\includegraphics[width=0.3\textwidth]{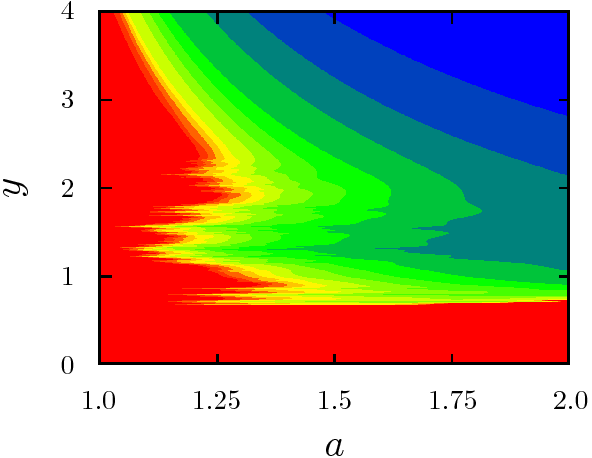}
\rule{4mm}{0mm}
\raisebox{31mm}{(c)}
\rule{-3mm}{0mm}
\includegraphics[width=0.3\textwidth]{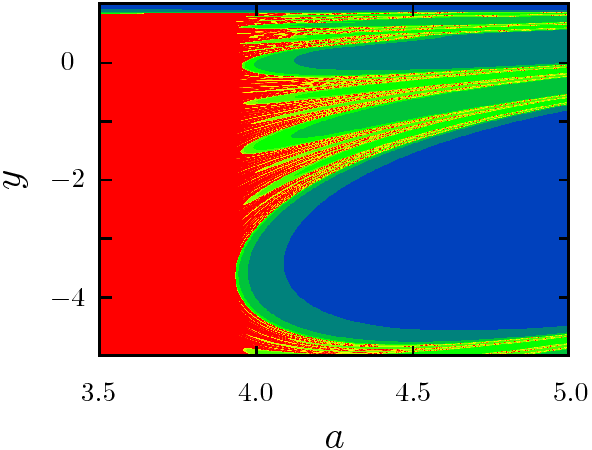}
}
\caption[]{\label{fig:parameter_plot}
  Metamorphosis of the $x=0$ section through the boundary for fixed 
  $\gamma =3.0$, $\epsilon=0.03$, and varying $\parX$. Panel (a) gives
  an overview, (b) a blow-up of the region with a rough basin boundary, 
  and (c) focusses on the transition of the chaotic attractor to a chaotic
  saddle. The colour coding is the same as in \Fig{rough_border};
  vertical sections through the present plots exactly agree with a
  section along the $y$ axis of the corresponding plot in
  \Fig{rough_border}. 
}
\end{figure}

To gain insight into the relation between these two kinds of lifetime
plots we first consider the conceptually simplest case where the
lifetime of trajectories starting on the $y$ axis is plotted as a
function of $\parX$ and $y$ (\Fig{parameter_plot}).
The large blue domain in the upper half indicates parameters and
initial conditions that are quickly attracted to the fixed point.  The
large red region in the lower left indicates initial conditions which
never get to the laminar fixed point, since the turbulent domain is an
attractor. 

The magnification \Fig{parameter_plot}(b) focusses on the fuzzy regions in
the lifetime plot for $\parX\simeq 1.25$. As we have seen in
\Fig{rough_border} the boundary between the two coexisting attractors in
the coordinate space $(x,y)$ is rough for these parameters.  As a
consequence the $y$-axis repeatedly crosses the boundary between the
domains of attraction of the respective attractors. This gives rise to
the observed spiky structure of the interface in the $\parX$-$y$ plot
\Fig{parameter_plot}(b).
Beyond $\parX\simeq 1.6$ the basin boundary is smooth
[\Fig{rough_border}(c,d)], and also in an $\parX$-$y$-plot there is a
sharp boundary between the two domains. It is located close to 
$y\simeq 0.82$.

When the attractor undergoes the boundary crisis at $\parX_{cr}=3.93$
the fingers from \Fig{saddle} are visible also in the
$\parX$-$y$-plot. They form a hierarchical structure of regions that
are mapped into the crisis region and subsequently rapidly approach
the laminar state.  Note that, when sufficiently resolved, also in
this case all fingers extend to the critical parameter value
$\parX_{cr} \simeq 3.93$.

\subsection{Generic parameter-coordinate dependence of the lifetime}

\begin{figure}
\rule{10mm}{0mm}
\raisebox{42mm}{(a)}
\includegraphics[width=0.4\textwidth]{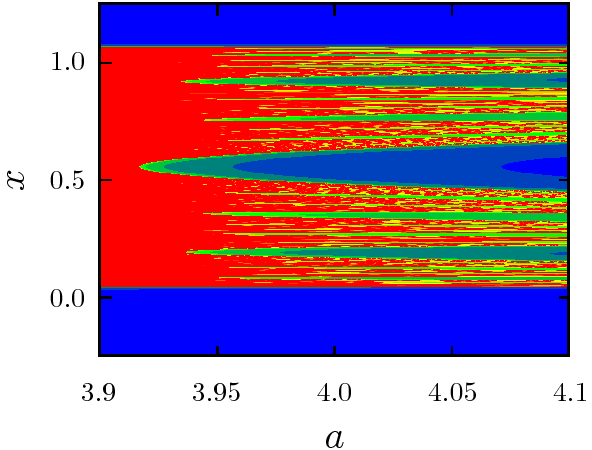}
\rule{10mm}{0mm}
\raisebox{42mm}{(b)}
\includegraphics[width=0.4\textwidth]{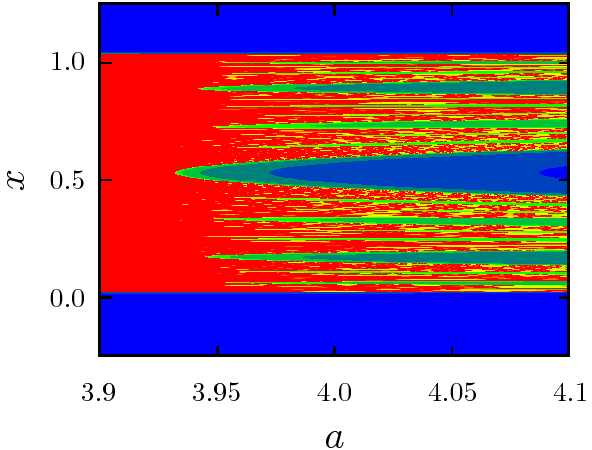}
\\[8mm]
\rule{10mm}{0mm}
\raisebox{42mm}{(c)}
\includegraphics[width=0.4\textwidth]{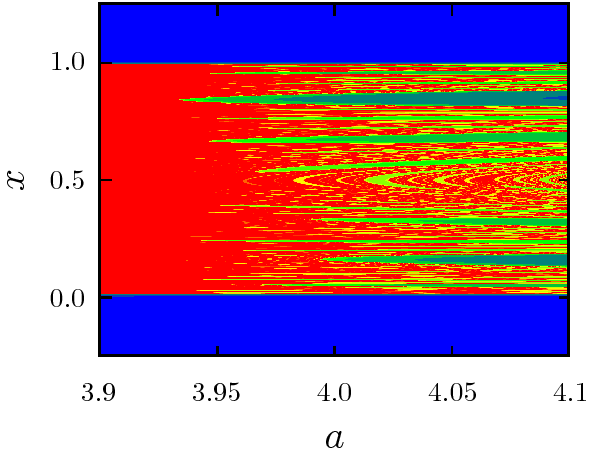}
\rule{10mm}{0mm}
\raisebox{42mm}{(d)}
\includegraphics[width=0.4\textwidth]{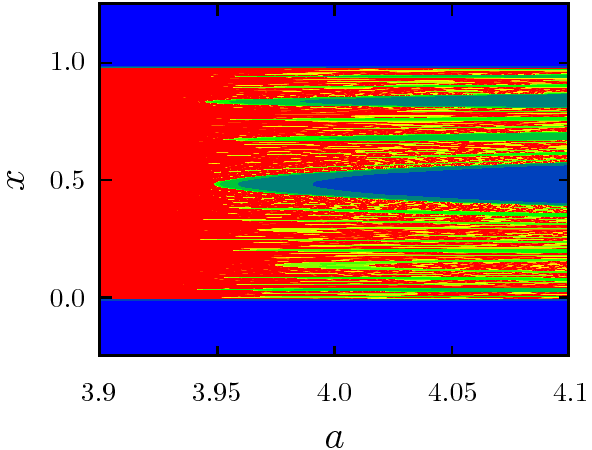}
\caption[]{\label{fig:fractal_islands}
  Metamorphosis of a section through the boundary 
  where the coordinates are varied along a line \eq{section}.
  The different panels correspond to (a) $m=1/3$,
  (b) $m=0$, (c) $m=-1/3$, and (d) $m=-2/3$, respectively. All other parameters
  and the colour coding are the same as in  \Fig{parameter_plot}. 
}
\end{figure}

In \Fig{parameter_plot} we chose 
a section aligned almost parallel to the edge of chaos.
On the other hand, in the applications \cite{darbyshire95,Sku06,Sch07a}
the amplitude of a perturbation of the laminar
state is varied, \ie initial conditions are chosen along a line
extending from the laminar fixed point towards the phase-space domain
admitting chaotic motion.
Such a line intersects the boundary more or less perpendicularly.
In that case one encounters a sharp, smoothly varying boundary between
the laminar and turbulent regions for all values below the crisis: It
is no longer possible to resolve the roughness of the boundary close to
$\parX \simeq 1.25$. In view of this we focus on the region close to
the crisis. The appropriate parts of the parameter plots for four
different slopes $m$ of the line 
\begin{equation}
  y  =  2 + m \; (x + 2)
\label{eq:section}
\end{equation}
are shown in \Fig{fractal_islands}. 

All panels of \Fig{fractal_islands} show hierarchical organised traces
of the fingers that we also saw in \Fig{parameter_plot}(c). This shows
that folded and hierarchically organised structures in lifetime-plots
are generic. They do not dependent on the specific direction along
which initial conditions are chosen. On the other hand the choices
differ in the detailed structure of the folds:
\FIG{fractal_islands}(a) shows the situation where the initial
conditions on the line approach the saddle, but do not
intersect it.
In this case the folds are nicely aligned, and they extend down to
different parameter values $\parX$ well below the bifurcation. After
all [\cf \Fig{saddle}(a)], the fingers invade the domain of attraction
before they collide with the attractor at the crisis, and the tips of
the finer fingers come down at a later time.
\FIG{fractal_islands}(b) corresponds to the situation where the
line touches the outer edge of the saddle. Consequently, it is exactly
along this line that all finger tips \emph{simultaneously} collide
with the attractor. Before the crisis, all initial conditions proceed
into the attractor, and at the crisis there is a fractal set of folds
with initial conditions escaping to the laminar state appearing all at
once. Subsequently, only the scaling of the width of the folds, and
hence the fractal dimension of the remaining saddle changes.
In \Fig{fractal_islands}(c) the initial conditions giving rise to
chaotic motion lie right in the heart of the chaotic invariant set. In
this case the folds also appear simultaneously at the crisis. A new
feature is that the internal dynamics of the saddle gives rise to a
non-trivial bending of the folds. For many values of $x$, in
particular $x=0.5$, there is not a unique value of $\parX$ which
separates regions of persistent chaotic motion for smaller $\parX$
from a decay to the laminar state. Rather, there can be multiple switching
between these possibilities as $\parX$ is increased.
When the line \eq{section} intersects the chaotic set only at
its lower boundary [\Fig{fractal_islands}(d)] the qualitative features
of the position-parameter plot are the same as in case (c), except
that the multiple switching is less pronounced.

In all cases the observed structure of folded hierarchical tongues are
reminiscent of the observations in studies of minimal perturbation
amplitudes in pipe flow \cite{darbyshire95,Sch07a}. Thus, this model
provides further support for the idea that transient turbulent motion
is generated by a chaotic saddle that coexists with a laminar fixed
point in the state space of linearly stable shear flows.
The following section discusses in more detail the implications of
these findings to the transition to turbulence in linearly stable
shear flows.

\section{Discussion}
\label{sec:turbulence}

\subsection{Methods}

We have suggested a low-dimensional model in which we can analyse
methods and concepts that recently been used in the framework of
fluid-mechanical systems
\cite{TohItano1999,Sku06,Sch07a,SchneiderGibson2007,DuguetWillisKerswell2007,Viswanath2007,ViswanathCvitanovic2008}.
The familiar concept of basin boundaries that separate different
attractors was extended to the situation of a saddle coexisting with
an attractor. We showed that the orbits defining the basin boundary
are a set that changes smoothly when crossing a crisis point where one
of the attractors looses its stability.  Beyond crisis we denote the
closure of this set as the \emph{edge of chaos}.  The edge can be
tracked by an iterative algorithm that exploits local properties only
and hence can be used both in the situation of co-existing attractors
as well as transient chaos coexisting with an attractor, see
\Fig{rel_attractor}(a,right panel).

A standard procedure to determine the basin boundary is
backward-iteration. It is more efficient than the direct forward
sampling of phase space which was used to generate \Fig{rough_border}.
The effort of backward iteration to determine a boundary of
box-counting dimension $D_B$ with a resolution $\epsilon$ scales like
$\epsilon^{-D_B}$. In contrast, the direct iteration scales
quadratically with the resolution, \ie like $\varepsilon^{-2}$.
The edge tracking algorithm adopted in the present work
(\Fig{rel_attractor}) roughly requires the same numerical effort as
backward iteration of the boundary of the region, and it has the
additional benefit that beyond the crisis it focusses on the
dynamically most relevant region of the edge of chaos while the
backward iteration also tracks the circumference of all fingers shown
in \Fig{saddle}.

\subsection{Geometry of the boundary}

The geometry of the boundary separating laminar and turbulent dynamics
can be studied in lifetime plots, where lifetime of initial conditions
is either analysed for fixed parameters as a function of state-space
coordinates, or by varying a parameter and a coordinate.

For fixed parameters the separating boundary can be smooth or rough.
The analysis in \sect{roughTransition} shows that roughness can be
observed only if (a)~the dynamics in the edge is chaotic, and (b)~the
Lyapunov exponent characterising the chaotic dynamics \emph{in} the
boundary is larger than the one in perpendicular direction. Roughness
of the boundary hence is an indicator that there is a strong
chaotic dynamics in the basin boundary. 
Since there is no a priori reason why the Lyapunov exponent pointing
out of the separating boundary should be large, it will be interesting
to identify a fluid mechanical realization of rough basin boundaries.
Ideally, the system should have a control parameter that influences
the ratio of the Lyapunov exponents in the longitudinal and transverse
directions. A good candidate might be Taylor-Couette flow between
independently rotating cylinders with a narrow gap, in which case
it is close to the planar shear flows mentioned earlier
\cite{Faisst}. But it might also be possible to find evidence
for rough boundaries in other parameter regions and geometries
where a multitude of attractors can coexist
\cite{Abshagen2005}.

We have shown here how features of the boundary in the phase space
relate to features in the parameter-coordinate space.
The latter representation is typically studied in hydrodynamic
systems where the Reynolds number Re is adopted as parameter.
Increasing Re the boundary shows folded hierarchical organised
tongue-like structures. In our model they appear shortly before or at
the parameters of the boundary crisis of the turbulent attractor. The
tongues have thus been related to the emergence of dynamical
connections \citeaffixed{RempelChianMacauRosa2004a}{\cf} between the
relative attractor on the edge of chaos and the attractor mimicking
stable turbulent motion.
These fingers result from the chaotic motion of the attractor
undergoing a crisis. The presence of similar tongue-like structures in
linearly stable shear flows \cite{darbyshire95,Moe04a,Moe04,Sch07a}
further supports the idea of a turbulence generating chaotic saddle in
these flows. The long persistence of turbulent motion, \ie its tiny
decay rate, may then be interpreted as another manifestation of
supertransients \cite{LaiWinslow1995,BrebanNusse2006}.

The local attractor embedded in the separating boundary -- the
\emph{edge state} is an object both of theoretical and practical
interest. 
 The model shows that the local attractor can be a fixed
point, a periodic orbit or a chaotic set. The type of dynamics in the
boundary can be chosen independently of whether turbulent motion is
generated by an attractor or a saddle. Thus, it is not a priori clear
which type of edge state one should expect in transitional shear
flows.  A chaotic edge state has been identified in pipe flow
\cite{Sch07a}, and a simple fixed point in plane Couette flow
\cite{SchneiderGibson2007}.  However, based on our present model we
expect that other flow geometries show edge states with various other
types of dynamics.

\subsection{Outlook}

The iterated edge tracking algorithm can be used to analyse any
dynamical system showing two coexisting types of dynamics \cite{cassak2007}. 
Without
additional input the method can be used to analyse the position of the
boundary and of trajectories in the boundary.  A promising future
application might be in control strategies, where the edge tracking is
used to identify target states for chaos control \cite{Schuster1999}.
In various technological applications one is interested to
intentionally induce turbulence or keep the flow laminar
\cite{Bewley2001,Hoegberg2003,Kawahara2005,Fransson2006,WangGibsonWaleffe2007}.
Up to now the setting up of the required effective control mechanisms
mostly relies on empirical strategies, long-term experience and
intuition. The edge tracking mechanism can provide additional
guidance by identifying flow structures on which actuators could
focus.

\subsection{Closing remarks}

The concept of the \emph{edge of chaos} 
provides a powerful framework to
analyse nonlinear dynamical systems where attractors coexist with a
chaotic saddle and where 
the traditional concept of basin boundaries can no longer be applied.
The approach still works for systems with several positive Lyapunov
exponents. In that situation it provides insight into local attractors
in the edge of chaos.

\ack

The authors acknowledge financial support from the Deutsche Forschungsgemeinschaft. 
They are grateful to Jeff Moehlis and Tam\'as T\'el for comments on
the manuscript.  J.V. also acknowledges discussions with Predrag
Cvitanovic and Bj\"orn Hof.


\end{document}